\documentclass[a4paper,12pt]{article}
\usepackage{cmap}
\usepackage[T2A,T1]{fontenc}
\usepackage[utf8]{inputenc}
\usepackage[english]{babel}

\usepackage{amsmath, amssymb}

\usepackage{indentfirst}

\usepackage[sort,compress]{cite}

\usepackage{graphicx}
\usepackage[margin=0.5em]{subcaption}

\usepackage{hyperref}

\DeclareMathOperator{\diag}{diag}
\DeclareMathOperator{\tr}{tr}
\DeclareMathOperator{\Expect}{\mathsf{E}}
\newlength{\subfigsize}
\title{On the possibility of a considerable quality improvement of a quantum ghost image by sensing in the object arm}
\author{D.\,A.\,Balakin \and A.\,V.\,Belinsky}
\date{\small Department of Mathematical Modeling and Computer Science,
Faculty of Physics, M.~V.~Lomonosov Moscow State University, Moscow 119991, Russia.}

\begin{document}

\selectlanguage{english}

\maketitle

\begin{abstract}
We consider a modification of the classical ghost imaging scheme
where an image of the research object is formed and acquired in the object arm.
It is used alongside the ghost image
to produce an estimate of the transmittance distribution of the object.
It is shown that it allows to weaken the deterioration of image quality
caused by nonunit quantum efficiency of the detectors,
including the case when the quantum image obtained in the object arm
is additionally affected by the noise caused by photons
that have not interacted with the object.
\end{abstract}

\section*{Introduction}

One of the important arguments in favour of quantum ghost imaging is
its capability of producing the lighting conditions
that are the most benign for the research object,
when the illumination effect (sometimes irreversible) on the object is minimal
\cite{quantum_image}.
This is particularly important when illuminating living creatures,
e.\,g., by X-rays.

\begin{figure}
\centering
\includegraphics[width=0.9\linewidth]{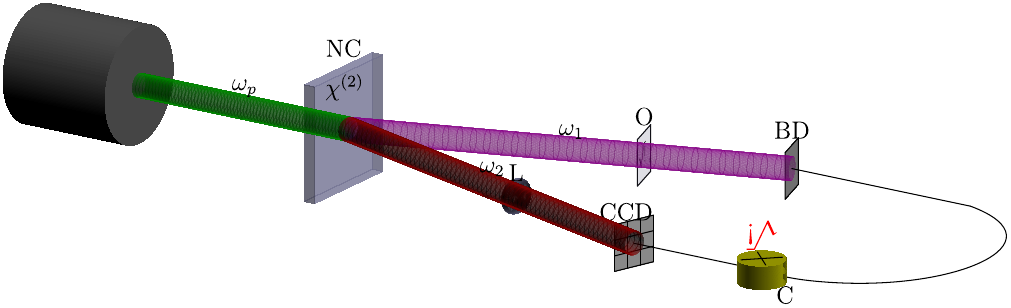}
\caption{Usual ghost imaging setup.
NC is a nonlinear crystal,
$\omega_p$ is pump radiation,
$\omega_1$ and $\omega_2$ are beams of entangled photon pairs
(the beams diverge due to use of noncollinear parametric scattering),
O is the object,
BD is the bucket detector in the object arm,
L is a converging lens,
CCD is a photodetector array in the reference arm,
C is an intensity correlator}
\label{fig:usual-gi}
\end{figure}

In usual ghost imaging
the sensing in the object arm (where the research object is)
is done by a bucket detector,
i.\,e. a single-pixel photodetector
intercepting the beam of the radiation
penetrating or reflected from the research object (see fig.~\ref{fig:usual-gi}).
Naturally, it does not produce an image.
An image is formed using the other arm and the correlations between photons
in these two arms.
For fixed optical setup
and given information about the research object available to the researcher
the minimal illumination of the object
at which acceptable image quality is achieved
can be lowered only by improving the quantum efficiency of the detectors.
However, the prospects of such improvement
are both limited and costly.
Moreover, while in ordinary imaging
it is only necessary to detect a photon that has interacted with the object,
ghost imaging requires detection of a \emph{pair} of photons,
thus, the mean number of detected photons
is proportional to \emph{squared} quantum efficiency.

\section{The proposed imaging setup}

We propose a new setup that provides,
on one hand, preservation of the studied object by lowering illumination intensity,
and on the other hand, improvement of image quality.
The improvement of the signal-to-noise ratio,
characteristic of ghost images formed by coincidence circuits, remains in force,
i.\,e. it is possible to combine the advantages of ghost imaging with the advantages of ordinary imaging.

\begin{figure}
\centering
\includegraphics[width=0.9\linewidth]{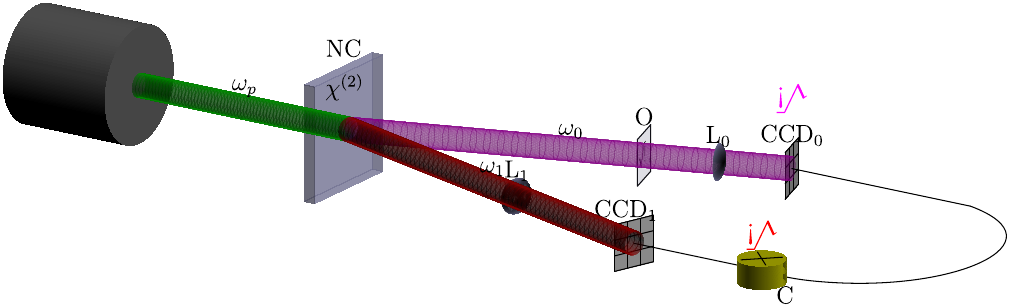}
\caption{The proposed setup to produce a pair of quantum images.
NC is a nonlinear crystal,
$\omega_p$ is pump radiation,
$\omega_1$ and $\omega_2$ are beams of entangled photon pairs
(the beams diverge due to use of noncollinear parametric scattering),
O is the object,
L$_0$ and L$_1$ are optical lenses,
CCD$_0$ and CCD$_1$ are photodetector arrays in object and reference arms,
C is an intensity correlator}
\label{fig:proposed-gi}
\end{figure}

Consider fig.~\ref{fig:proposed-gi}.
The object arm has a detector array instead of a bucket detector,
similarly to the reference arm.
An ordinary image is formed on the array
using an optical lens.
Alternatively, in the case of X-ray illumination
the array is placed immediately behind the object,
as only reflective optical systems with large incidence angles work well enough in the X-ray range.

Hence, in the proposed setup, two images, an ordinary one and a quantum one,
are produced.
Their computer processing allows to lower the minimum number of illumination photons
and to improve the image quality.
Note that the proposed setup differs from difference measurements
(see, e.\,g. \cite{treps_et_al_2005}),
as unlike them
the acquired images are processed not by taking their difference.
This means, in particular, that the requirement of identity of the sensors
in object and reference arms is lifted.
Furthermore, by virtue of radically different principles of our setup
and the setup of difference measurements,
the layouts of setups differs as well.

\section{Measurement reduction method}

The mathematical methods and algorithms for processing
image(s) of an object that is illuminated by a low number of photons
have to, in addition to providing maximal precision,
enable use of all information about the object that is available to the researcher.
The mathematical method of measurement reduction and its algorithmic implementations
allow to achieve this.

Consider a typical measurement setup,
in which the measured signal $f$ belonging to a Euclidean space $\mathcal{F}$,
see \cite{pytyev_ivs},
is formed at the input of a measuring transducer (MT).
The MT transforms $f$ into a signal
\begin{equation}
\label{eqn:measurement-scheme}
\xi = A f + \nu,
\end{equation}
that belongs to the Euclidean space $\mathcal{X}$,
where $A \colon \mathcal{F} \to \mathcal{X}$
is the operator that models physical processes in the MT
defining the transformation of $f$ into the signal $A f$
and denotes the modeled MT,
$\nu$ is the measurement error.
The measurement result depends on the characteristics of the measured object
that interacts with the MT and is distorted by the measurement,
while the researcher is usually interested in the characteristics of the object undistorted by the measurement.
Their relation is modeled by an ideal MT that is given by the operator $U \colon \mathcal{F} \to \mathcal{U}$.
The ideal MT transforms the same signal $f$ into the signal $U f$
representing the features of interest to the researcher of the research object.
The reduction problem consists of finding the reduction operator $R_*$
for which $R_* \xi$  is the most accurate version of $U f$.
If in \eqref{eqn:measurement-scheme} $f$ is an a priori arbitrary vector,
$\nu$ is a random vector taking values in $\mathcal{X}$
that has expectation $\Expect \nu = 0$ and a non-degenerate covariance operator
$\Sigma_{\nu}$: $\forall x \in \mathcal{X}$ $\Sigma_{\nu} x = \Expect \nu (x, \nu)$,
the linear reduction operator $R_* \colon \mathcal{X} \to \mathcal{U}$ is defined as the one minimizing the worst-case (in $f$) mean squared error (MSE) of interpreting $R \xi$ as $U f$:
\begin{equation*}
h(R, U) = \sup_{f \in \mathcal{F}} \lVert R \xi - U f\rVert^2.
\end{equation*}
The MSE is minimal \cite{pytyev_ivs} for
\begin{equation}
\label{eqn:lin-red-op}
R_* = U (A^* \Sigma_{\nu}^{-1} A)^{-1} A^* \Sigma_{\nu}^{-1}
\end{equation}
and is equal to
\begin{equation}
\label{eqn:lin-red-mse}
h(R_*, U) = \tr U (A^* \Sigma_{\nu}^{-1} A)^{-1} U^*,
\end{equation}
where ${}^-$ denotes pseudoinversion (Moore--Penrose inversion),
if $U (I - A^- A) = 0$, and is infinite otherwise.

Suppose that the researcher is interested in the transmittance distribution of the object.
The values of pixel transmittances belong to the unit range.
This is taken into account during measurement reduction
by projecting the estimate onto the set $[0, 1]^{\dim \mathcal{U}}$
\cite{reduction_vmu, reduction_uzmu, ghost_images_jetp}:
the estimate is the fixed point of the mapping
that combines the linear reduction result $R_* \xi$
and some estimate $\hat{u}$ of the transmittance distribution
as uncorrelated results of the ``main'' and dummy measurements
with subsequent projection onto $[0, 1]^{\dim \mathcal{U}}$
while the minimized distance is the Mahalanobis distance
$\lVert \Sigma_{R_* \xi}^{-1/2} \cdot \rVert$
that is induced by the covariance operator
$\Sigma_{R_* \xi} = U (A^* \Sigma_{\nu}^{-1} A)^{-1} U^*$
of the linear reduction estimate.
In addition, the researcher knows that transmittances of neighbouring pixels usually do not differ much.
This information is formalized
\cite{imaging_small_n_photons, gi_compressed_sensing_substr_const, hi-res_gi_sparsity, compressive_ghost_imaging, katz_et_al}
by \emph{sparisty} of the transmittance distribution as a vector $U f$ in a given basis,
i.\,e. as the information that in this basis
most components of $U f$ are zero.
In \cite{mgi_vmu, balakin_et_al_2019}, an algorithm of measurement reduction
that allows to take such information into account
when processing multiplexed quantum ghost images was proposed.
The algorithm is based on testing statistical hypotheses that
estimate components in the chosen basis
are zero
(alternative hypotheses are that the components are non-zero).
Testing results depend on the algorithm parameter $\tau$
that is related to the significance level of the used criterion.
Its choice is determined by the compromise between noise suppression and image distortion
that is acceptable to the researcher.
Distortion is understood as a \emph{deterministic} (as opposed to the noise) change
during processing
that brings the image away from the ideal one.

\section{Effects of detector quantum efficiency on the measurement setup}

Suppose that CCD arrays in object and reference arms are identical and are described by the operator $A_0$.
In the case of unit quantum efficiency of the detectors the measurement setup \eqref{eqn:measurement-scheme} is as follows:
\begin{equation*}
\xi =
\begin{pmatrix}
\xi_0\\
\xi_1
\end{pmatrix}
=
\begin{pmatrix}
A_0\\
A_0
\end{pmatrix}
n f
+
\begin{pmatrix}
\nu_{\textnormal{img}} + \nu_{\epsilon}\\
\nu_{\textnormal{img}}
\end{pmatrix},
\end{equation*}
where $\xi_0$ is the result of the measurement by the CCD array in the object arm,
$\xi_{1}$ is the measured obtained ghost image,
$\nu_{\textnormal{img}}$ is the part of error caused by quantum image acquisition,
$\nu_\epsilon$ is the part of error of the quantum image acquired in the object arm
that is caused by noise photons
(during ghost imaging they are mostly suppressed by the coincidence circuit
if its trigger time is sufficiently small,
see comparison of quantum imaging setups in \cite{imaging_small_n_photons};
otherwise a similar term appears in the error of the acquired quantum ghost image),
$n$ is the mean number of photons illuminating the object.
The error covariance operator has the form
\begin{equation}
\label{eqn:cov-op-unit-eff}
\Sigma_{\nu} =
\begin{pmatrix}
A_0 S(f) A_0^* + \Sigma_{\nu_{\epsilon}} & A_0 S(f) A_0^*\\
A_0 S(f) A_0^* & A_0 S(f) A_0^*
\end{pmatrix}
=
\begin{pmatrix}
1 & 1\\
1 & 1\\
\end{pmatrix}
\otimes
A_0 S(f) A_0^*
+
\begin{pmatrix}
\Sigma_{\nu_{\epsilon}} & 0\\
0 & 0
\end{pmatrix}
\end{equation}
where the operator $S(f)$ is determined by photocount variances and covariances.
For example, if photocounts are Poissonian,
$S(f) = n \diag(f)$.
Here and below $\diag(q)$ denotes the matrix
whose diagonal elements are equal to corresponding components of $q$
and offdiagonal elements are zero,
and $\otimes$ denotes Kronecker product.
In this case acquisition of spatial intensity distribution in the object arm
does not allow to improve reduction MSE
(the reduction result does not depend on the result of the measurement in the object arm).

For object arm detector efficiency $\eta_0$
and reference arm detector efficiency $\eta_1$
the measurement setup becomes
\begin{equation*}
\xi =
\begin{pmatrix}
\xi_0\\
\xi_1
\end{pmatrix}
=
\begin{pmatrix}
\eta_0 A_0\\
\eta_0 \eta_1 A_0
\end{pmatrix}
n f
+
\begin{pmatrix}
\nu_{\textnormal{img}} + \nu_0 + \nu_{\epsilon}\\
\nu_{\textnormal{img}} + \nu_0 + \nu_1
\end{pmatrix},
\end{equation*}
where the additional errors $\nu_0$ and $\nu_1$
are caused by missing of some photons by the sensors in object and reference arms,
respectively.
The error covariance operator becomes
\begin{multline}
\label{eqn:cov-op}
\Sigma_{\nu} =
\eta_0^2
\begin{pmatrix}
1 & \eta_1\\
\eta_1 & \eta_1^2
\end{pmatrix}
\otimes
A_0 S(f) A_0^*
+\\+
\eta_0^2
\begin{pmatrix}
\Sigma_{\nu_{\epsilon}} & 0\\
0 & 0
\end{pmatrix}
+
\begin{pmatrix}
\eta_0 (1 - \eta_0) & \eta_0 (1 - \eta_0) \eta_1\\
\eta_0 (1 - \eta_0) \eta_1 & \eta_0 \eta_1 (1 - \eta_0 \eta_1)
\end{pmatrix}
\otimes
\diag(A_0 f),
\end{multline}
where the first two terms have the same origin as in \eqref{eqn:cov-op-unit-eff},
but are weakened by the decrease of the mean number of photocounts,
and the last term is directly related to non-unit detector efficiency.

\begin{figure}
\centering
\includegraphics[scale=1]{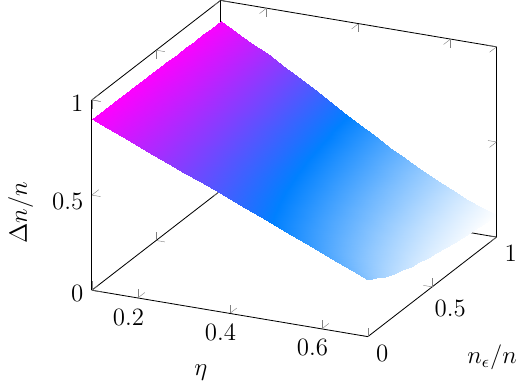}
\caption{Relative gain $\Delta n / n$ in the mean number of illumination photons
as a function of detector quantum efficiency $\eta = \eta_0 = \eta_1$
and the relative mean number of noise photons $n_\epsilon / n$
(relative to the mean number of illumination photons $n$,
when only the ghost image is acquired)}
\label{fig:mean-photon-number-gain}
\end{figure}

As the linear reduction operator has the form \eqref{eqn:lin-red-op},
while the reduction estimate MSE is equal to \eqref{eqn:lin-red-mse},
the gain caused by image acquisition in the object arm
is equal to
\begin{multline}
\label{eqn:mse-gain}
\eta_0^{-2}
\tr
U
\left(
\eta_0 \eta_1^{-1}
\left(
A_0^* \left(
\eta_0 \eta_1 A_0 S(f) A_0^* + (1 - \eta_0 \eta_1) \diag(A_0 f)
\right)^{-1} A_0
\right)^{-1}
\right.
-\\-
\left.
\left(
\begin{pmatrix}
A_0^* & \eta_1 A_0^*
\end{pmatrix} \Sigma_{\nu}^{-1} \begin{pmatrix}
A_0\\
\eta_1 A_0
\end{pmatrix}
\right)^{-1}
\right)
U^*,
\end{multline}
where the operator $\Sigma_{\nu}$ is defined by the expression \eqref{eqn:cov-op}.
Fig.~\ref{fig:mean-photon-number-gain} shows the relative gain in mean number of illumination photons
as a function of detector quantum efficiency and the relative mean number of noise photons
(relative to the mean number of illumination photons
when only the ghost image is acquired).
The gain is defined as the ration of the decrease $\Delta n$ in the mean number of photons
that is required to illuminate the object
to the mean number of illumination photons $n$ in the usual ghost imaging setup
so that estimate MSE remains the same.
It is expected that the number of noise photons decreases proportionally.
As expected, the gain increases monotonically
if the number of noise photons decreases
or detector quantum efficiency $\eta = \eta_0 = \eta_1$ decreases
(in the figure we assume that detector efficiencies in both arms are equal).
Without noise photons the relative gain is $1 - \eta$.
If additional information is available,
the gain usually decreases while preserving its sign,
as both compared MSEs decrease,
but the greater one decreases more.

\section{Computer modeling results}

\begin{figure}
\centering
\begin{subfigure}[t]{\subfigsize}
\centering
\includegraphics[scale=1]{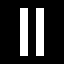}
\caption{Research object transmittance distribution}
\label{fig:fig3_src}
\end{subfigure}
\begin{subfigure}[t]{\subfigsize}
\centering
\includegraphics[scale=1]{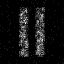}
\caption{The image acquired by the CCD array in the object arm}
\label{fig:fig3_obj}
\end{subfigure}
\begin{subfigure}[t]{\subfigsize}
\centering
\includegraphics[scale=1]{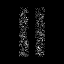}
\caption{The ghost image (GI)}
\label{fig:fig3_ref}
\end{subfigure}
\begin{subfigure}[t]{\subfigsize}
\centering
\includegraphics[scale=1]{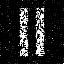}
\caption{Reduction, no sparsity information}
\label{fig:fig3_red}
\end{subfigure}
\begin{subfigure}[t]{\subfigsize}
\centering
\includegraphics[scale=1]{{{fig3_red-0.1}}}
\caption{Reduction with sparsity information, $\tau = 0.1$}
\label{fig:fig3_red-0}
\end{subfigure}
\begin{subfigure}[t]{\subfigsize}
\centering
\includegraphics[scale=1]{{{fig3_red-0.2}}}
\caption{Reduction with sparsity information, $\tau = 0.2$}
\label{fig:fig3_red-1}
\end{subfigure}
\begin{subfigure}[t]{\subfigsize}
\centering
\includegraphics[scale=1]{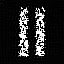}
\caption{GI reduction, no sparsity information}
\label{fig:fig3_red-s}
\end{subfigure}
\begin{subfigure}[t]{\subfigsize}
\centering
\includegraphics[scale=1]{{{fig3_red-0.1-s}}}
\caption{GI reduction with sparsity information, $\tau = 0.1$}
\label{fig:fig3_red-0-s}
\end{subfigure}
\begin{subfigure}[t]{\subfigsize}
\centering
\includegraphics[scale=1]{{{fig3_red-0.2-s}}}
\caption{GI reduction with sparsity information, $\tau = 0.2$}
\label{fig:fig3_red-1-s}
\end{subfigure}
\caption{Quantum images of (\subref{fig:fig3_src})~the object
that were acquired in  (\subref{fig:fig3_obj})~object and (\subref{fig:fig3_ref})~reference arms
and the results of their processing using measurement reduction technique:
the results of processing both images
(\subref{fig:fig3_red})~without using sparsity information
and (\subref{fig:fig3_red-0}, \subref{fig:fig3_red-1})~with using sparsity information for two values of the algorithm parameter $\tau$;
(\subref{fig:fig3_red-s}--\subref{fig:fig3_red-1-s})~the results of processing only the ghost image using the same methods}
\label{fig:fig3}
\end{figure}

Fig.~\ref{fig:fig3} shows the results of computer modeling and processing of quantum images using measurement reduction method via the algorithm described in \cite{mgi_vmu, balakin_et_al_2019}.
For modeling it was assumed that the CCD arrays in the arms are identical,
the sensors are three times as large as an object pixel,
each object pixel is illuminated by an average of $1$ photon,
the mean number of noise photons that did not interact with the object
but were detected by the sensor in the object arm is $0.1$ per pixel
and detector quantum efficiencies are $0.4$.

As noted above, the algorithm parameter $\tau$
reflects a compromise between noise suppression and image distortion
that is acceptable for the researcher.
The larger the value of $\tau$, the larger noise suppression,
but the distortion of image details increases as well
(cf., e.\,g., figs.~\ref{fig:fig3_red-0-s} and \ref{fig:fig3_red-1-s},
where increasing $\tau$ makes slit image mored blurred).
The value $\tau = 0$ corresponds to both
absence of additional noise suppression
using sparsity information about the transmittance distribution
and absence of image distortion,
i.\,e. it is equivalent to lack of information about the sparsity of object transmittance distribution.
The values of $\tau$ shown in fig.~\ref{fig:fig3}
are chosen so that in fig.~\ref{fig:fig3_red-0} the distortions are not be noticeable yet,
and in fig.~\ref{fig:fig3_red-1} they are be noticeable.

The above suggestion that
if additional information is available,
the gain caused by processing the image acquired in the object arm
decreases while preserving its sign
is supported by the comparison of pairs of figs.~\ref{fig:fig3_red} and \subref{fig:fig3_red-s}
and figs.~\ref{fig:fig3_red-0} and \subref{fig:fig3_red-0-s}:
the former two differ more than the latter two,
as the set of possible estimate values is restricted.

The estimate built on only the ghost image
is more noisy than the estimate build on both images.
Therefore, the blurring for the same values of the algorithm parameter $\tau$
that is responsible for the balance between
blurring of important image details and noise suppression
is larger for the estimate that uses only the ghost image.

Finally, the localization of the noise caused by the image acquired in the object arm
differs from the localization of the noise caused by non-unit detector quantum efficiency.
While the spatial distribution of noise caused by noise photons usually is independent of the object,
the distribution of the noise caused by detector inefficiency is directly related to it,
since the larger the number of the arrived photons, the larger the variance of the number of photocounts.
For that reason the change of the parameter $\tau$ affects distortion of image details differently for areas with different average brightness.

\section*{Conclusion}

In conclusion, we note that the considered imaging setup allows to improve the quantum image quality
during computer processing step
also by taking diffraction limits of ghost imaging into consideration in this step \cite{belinsky_2018, moreau_et_al_2018, moreau_et_al_2018a}.
This is caused by the absence of diffraction limits on ordinary imaging
that are inherent in ghost imaging
and are related to limits on transversal pump size and, consequently, its divergence.

In addition, we would like to note an interesting setup
based on the proposed one:
simultaneous illumination of two objects that are located in different arms.
In this way, 2 pairs of images (2 ordinary images and 2 ghost images)
are produced at the same time
at the cost of worsening the quality compared to absence of the second object.
The cause of worsening is that the second object does not transmit some of the photons
that are correlated with photons in the arm of the first object
and vice versa.
Nevertheless, this could be useful in some situations.

This work was supported by the Russian Foundation for Basic Research, project no.~18-01-00598~A.

\bibliographystyle{ugost2008}
\newcommand{\BibDash}{}
\bibliography{gi-impr-obj-eng}

\begin{thebibliography}{10}
\def\selectlanguageifdefined#1{
\expandafter\ifx\csname date#1\endcsname\relax
\else\selectlanguage{#1}\fi}
\providecommand*{\href}[2]{{\small #2}}
\providecommand*{\url}[1]{{\small #1}}
\providecommand*{\BibUrl}[1]{\url{#1}}
\providecommand{\BibAnnote}[1]{}
\providecommand*{\BibEmph}[1]{#1}
\ProvideTextCommandDefault{\cyrdash}{\iflanguage{russian}{\hbox
  to.8em{--\hss--}}{\textemdash}}
\providecommand*{\BibDash}{\ifdim\lastskip>0pt\unskip\nobreak\hskip.2em plus
  0.1em\fi
\cyrdash\hskip.2em plus 0.1em\ignorespaces}
\renewcommand{\newblock}{\ignorespaces}

\bibitem{quantum_image}
\selectlanguageifdefined{english}
\href{http://dx.doi.org/10.1007/0-387-33988-4}{Quantum imaging}~/ Ed.\ by\
  M.~I.~Kolobov. \BibDash
\newblock Springer New York, 2007.

\bibitem{treps_et_al_2005}
\selectlanguageifdefined{english}
Quantum noise in multipixel image processing~/ N.~Treps, V.~Delaubert,
  A.~Ma{\^{i}}tre et~al.~//
  \href{http://dx.doi.org/10.1103/physreva.71.013820}{\BibEmph{Physical Review
  A}}. \BibDash
\newblock 2005. \BibDash
\newblock Vol.~71, no.~1. \BibDash
\newblock P.~013820.

\bibitem{pytyev_ivs}
\selectlanguageifdefined{english}
\BibEmph{Pyt'ev~Yu.~P.} Methods of mathematical modeling of
  measuring--computing systems [in {R}ussian]. \BibDash
\newblock 3 edition. \BibDash
\newblock Moscow~: Fizmatlit, 2012.

\bibitem{reduction_vmu}
\selectlanguageifdefined{english}
\BibEmph{Balakin~D.~A., Pyt'ev~Yu.~P.} A comparative analysis of reduction
  quality for probabilistic and possibilistic measurement models~//
  \href{http://dx.doi.org/10.3103/s0027134917020047}{\BibEmph{Moscow University
  Physics Bulletin}}. \BibDash
\newblock 2017. \BibDash
\newblock Vol.~72, no.~2. \BibDash
\newblock P.~101--112.

\bibitem{reduction_uzmu}
\selectlanguageifdefined{english}
\BibEmph{Balakin~D.~A., Pyt'ev~Yu.~P.} Improvement of measurement reduction if
  the feature of interest of the research object belongs to a known convex
  closed set [in Russian]~// \BibEmph{Memoirs of the Faculty of Physics}.
  \BibDash
\newblock 2018. \BibDash
\newblock no.~5. \BibDash
\newblock P.~1850301.

\bibitem{ghost_images_jetp}
\selectlanguageifdefined{english}
\BibEmph{Balakin~D.~A., Belinsky~A.~V., Chirkin~A.~S.} Improvement of the
  optical image reconstruction based on multiplexed quantum ghost images~//
  \href{http://dx.doi.org/10.1134/S1063776117070147}{\BibEmph{Journal of
  Experimental and Theoretical Physics}}. \BibDash
\newblock 2017. \BibDash
\newblock Vol. 125, no.~2. \BibDash
\newblock P.~210--222.

\bibitem{imaging_small_n_photons}
\selectlanguageifdefined{english}
Imaging with a small number of photons~/ P.~A.~Morris, R.~S.~Aspden,
  J.~E.~C.~Bell et~al.~//
  \href{http://dx.doi.org/10.1038/ncomms6913}{\BibEmph{Nat. Commun.}} \BibDash
\newblock 2015. \BibDash
\newblock Vol.~6. \BibDash
\newblock P.~5913.

\bibitem{gi_compressed_sensing_substr_const}
\selectlanguageifdefined{english}
Image quality enhancement in low-light-level ghost imaging using modified
  compressive sensing method~/ X.~Shi, X.~Huang, S.~Nan et~al.~//
  \href{http://dx.doi.org/10.1088/1612-202x/aaa5f6}{\BibEmph{Laser Phys.
  Lett.}} \BibDash
\newblock 2018. \BibDash
\newblock Vol.~15, no.~4. \BibDash
\newblock P.~045204.

\bibitem{hi-res_gi_sparsity}
\selectlanguageifdefined{english}
\BibEmph{Gong~W., Han~S.} High-resolution far-field ghost imaging via sparsity
  constraint~// \href{http://dx.doi.org/10.1038/srep09280}{\BibEmph{Sci. Rep.}}
  \BibDash
\newblock 2015. \BibDash
\newblock Vol.~5, no.~1. \BibDash
\newblock P.~9280.

\bibitem{compressive_ghost_imaging}
\selectlanguageifdefined{english}
Entangled-photon compressive ghost imaging~/ P.~Zerom, K.~W.~C.~Chan,
  J.~C.~Howell, R.~W.~Boyd~//
  \href{http://dx.doi.org/10.1103/physreva.84.061804}{\BibEmph{Phys. Rev. A}}.
  \BibDash
\newblock 2011. \BibDash
\newblock Vol.~84, no.~6. \BibDash
\newblock P.~061804.

\bibitem{katz_et_al}
\selectlanguageifdefined{english}
\BibEmph{Katz~O., Bromberg~Y., Silberberg~Y.} Compressive ghost imaging~//
  \href{http://dx.doi.org/10.1063/1.3238296}{\BibEmph{Appl. Phys. Lett.}}
  \BibDash
\newblock 2009. \BibDash
\newblock Vol.~95, no.~13. \BibDash
\newblock P.~131110.

\bibitem{mgi_vmu}
\selectlanguageifdefined{english}
\BibEmph{Balakin~D.~A., Belinsky~A.~V.} Reduction of multiplexed quantum ghost
  images~// \href{http://dx.doi.org/10.3103/S0027134919010053}{\BibEmph{Moscow
  University Physics Bulletin}}. \BibDash
\newblock 2019. \BibDash
\newblock Vol.~74, no.~1. \BibDash
\newblock P.~8--16.

\bibitem{balakin_et_al_2019}
\selectlanguageifdefined{english}
\BibEmph{Balakin~D.~A., Belinsky~A.~V., Chirkin~A.~S.} Object reconstruction
  from multiplexed quantum ghost images using reduction technique~//
  \href{http://dx.doi.org/10.1007/s11128-019-2193-x}{\BibEmph{Quantum
  Information Processing}}. \BibDash
\newblock 2019. \BibDash
\newblock Vol.~18, no.~3. \BibDash
\newblock P.~80.

\bibitem{belinsky_2018}
\selectlanguageifdefined{english}
\BibEmph{Belinsky~A.~V.} The ``paradox'' of {K}arl {P}opper and its connection
  with the heisenberg uncertainty principle and quantum ghost images~//
  \href{http://dx.doi.org/10.3103/S0027134918050053}{\BibEmph{Moscow University
  Physics Bulletin}}. \BibDash
\newblock 2018. \BibDash
\newblock Vol.~73, no.~5. \BibDash
\newblock P.~447--456.

\bibitem{moreau_et_al_2018}
\selectlanguageifdefined{english}
Experimental Limits of Ghost Diffraction: Popper's Thought Experiment~/
  P.-A.~Moreau, P.~A.~Morris, E.~Toninelli et~al.~//
  \href{http://dx.doi.org/10.1038/s41598-018-31429-y}{\BibEmph{Scientific
  Reports}}. \BibDash
\newblock 2018. \BibDash
\newblock Vol.~8, no.~1. \BibDash
\newblock P.~13183.

\bibitem{moreau_et_al_2018a}
\selectlanguageifdefined{english}
Resolution limits of quantum ghost imaging~/ P.-A.~Moreau, E.~Toninelli,
  P.~A.~Morris et~al.~//
  \href{http://dx.doi.org/10.1364/oe.26.007528}{\BibEmph{Optics Express}}.
  \BibDash
\newblock 2018. \BibDash
\newblock Vol.~26, no.~6. \BibDash
\newblock P.~7528.

\end{thebibliography}

\end{document}


\maketitle

\begin{abstract}
Рассмотрена модификация классической схемы формирования квантовых фантомных изображений,
в которой в объектном канале формируется и регистрируется
изображение исследуемого объекта.
Оно используется вместе с фантомным изображением
для построения оценки распределения прозрачности объекта.
Показано, что это позволяет ослабить ухудшение качества изображения,
связанное с неединичной квантовой эффективностью датчиков,
причем и в том случае, если квантовое изображение, полученное в объектном канале,
дополнительно поражено шумом, обусловленным фотонами,
не взаимодействовавшими с объектом.
\end{abstract}

\section*{Введение}

Одним из важных доводов в пользу использования квантовых фантомных изображений является
создание максимально щадящих условий освещения исследуемого объекта,
когда воздействие излучения на объект (иногда необратимое) минимально \cite{quantum_image}.
Особенно это важно при облучении живых существ, например, рентгеновским излучением.

\begin{figure}
\centering
\includegraphics[width=0.9\linewidth]{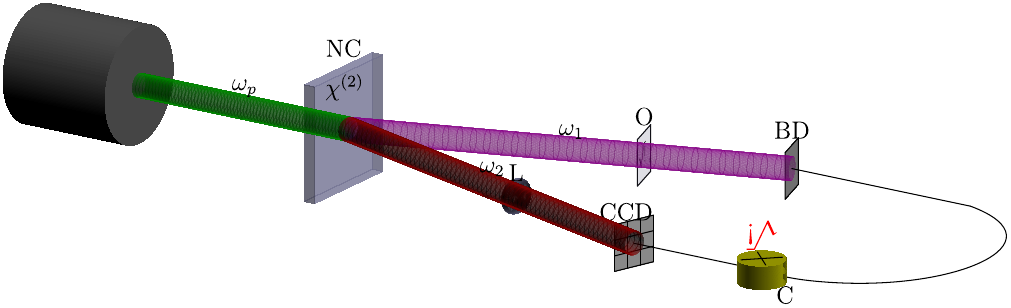}
\caption{Классическая схема формирования фантомных изображений.
NC~--- нелинейный кристалл;
$\omega_p$~--- накачка;
$\omega_1$ и $\omega_2$~--- пучки запутанных пар фотонов
(пучки расходятся вследствие использования неколлинеарного процесса
параметрического рассеяния);
O~--- объект;
BD~--- интегрирующий детектор в объектном канале;
L~--- собирающая линза;
CCD~--- матрица фотодетекторов в восстанавливающем канале;
C~--- коррелятор интенсивностей}
\label{fig:usual-gi}
\end{figure}

В классической схеме формирования фантомных изображений
в объектном канале (где находится исследуемый объект)
излучение регистрируется интегрирующим детектором,
т.\,е. одноэлементным фотоприемником,
перехватывающим весь пучок излучения,
пронизывающего или отраженного от объекта исследования (см. рис.~\ref{fig:usual-gi}).
При этом никакого изображения, конечно, не получается.
Оно формируется в другом канале
за счет жесткой корреляции между фотонами в этих двух каналах.
При фиксированных свойствах оптической схемы
и заданной информации об объекте исследования,
доступной исследователю,
минимальная освещенность объекта,
при которой достижимо приемлемое качество изображения,
может быть снижена лишь за счет
повышения квантовой эффективности детекторов.
Однако, резервы такого форсирования не только не беспредельны,
но и сопряжены с техническими трудностями и с финансовыми затратами.
При этом, если для обычных изображений
требуется лишь регистрация взаимодействовавшего с объектом фотона,
при формировании фантомных изображений требуется регистрация \emph{пары} фотонов,
т.\,е. среднее число зарегистрированных фотонов
пропорционально \emph{квадрату} квантовой эффективности.

Мы предлагаем новое схемное решение, обеспечивающее,
с одной стороны, сохранность объекта изучения за счет снижения интенсивности освещения,
а с другой,~--- повышение качества изображения.
При этом улучшение соотношения сигнал/шум,
характерное для фантомных изображений, формируемых схемой совпадений,
остается в силе,
т.\,е удается объединить преимущества фантомных изображений
с формированием обычных изображений.

\begin{figure}
\centering
\includegraphics[width=0.9\linewidth]{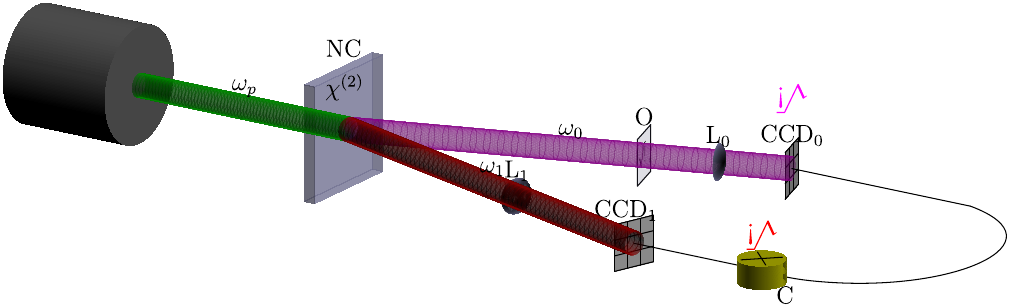}
\caption{Предлагаемая схема формирования пары квантовых изображений.
NC~--- нелинейный кристалл;
$\omega_p$~--- накачка;
$\omega_1$ и $\omega_2$~--- пучки запутанных пар фотонов
(пучки расходятся вследствие использования неколлинеарного процесса);
O~--- объект;
L$_0$, L$_1$~--- оптические объективы;
CCD$_0$, CCD$_1$~--- матрицы фотодетекторов в объектном и восстанавливающем каналах;
C~--- коррелятор интенсивностей}
\label{fig:proposed-gi}
\end{figure}

Рассмотрим рис.~\ref{fig:proposed-gi}.
В объектном канале вместо интегрирующего детектора~---
матрица фотоприемников,
как и в восстанавливающем канале,
на которой строится обыкновенное изображение исследуемого объекта
при помощи оптического объектива.
Либо, при облучении рентгеновским излучением,
матрица помещается непосредственно за объектом,
поскольку в рентгеновской области хорошо работают
лишь отражательные оптические системы с большими углами падения
и качество изображения оставляет желать лучшего.

Таким образом, в рассматриваемой схеме регистрируется два квантовых изображения:
обычное и фантомное.
Последующая их компьютерная обработка
дает возможность снизить минимально требуемое число фотонов, пронизывающих объект,
и повысить качество изображения.
Заметим, что предложенная схема отличается и от разностных измерений
(см., например, \cite{treps_et_al_2005}),
поскольку, в отличие от них,
полученные изображния обрабатываются не вычислением их разности.
Это означает, в частности, то, что снимается
жесткое условие абсолютной идентичности детекторов в объектном и восстанавливаюшем каналах.
Кроме того, в силу радикально различных принципов работы нашей и разностных схем,
отличается и компоновка их оптических систем.

\section{Метод редукции измерения}

В ситуации освещения объекта минимальным числом фотонов
эффективность математических методов и алгоритмов
обработки измерений должна быть такой,
чтобы они не только обеспечивали минимальную погрешность,
но и позволяли использовать всю доступную исследователю информацию об объекте.
Это позволяют сделать
математический метод редукции измерения
и реализующие его алгоритмы.

Рассмотрим типичную схему измерений,
в которой на входе измерительного преобразователя (ИП)
формируется измеряемый сигнал $f$,
принадлежащий евклидову пространству $\mathcal{F}$, см. \cite{pytyev_ivs}.
ИП преобразует $f$ в принадлежащий евклидову пространству $\mathcal{X}$ сигнал
\begin{equation}
\label{eqn:measurement-scheme}
\xi = A f + \nu,
\end{equation}
где $A \colon \mathcal{F} \to \mathcal{X}$~---
оператор, моделирующий физические процессы в ИП,
определяющие преобразование $f$ в сигнал $Af$,
и далее также обозначающий моделируемый им ИП,
$\nu$ --- погрешность, шум измерения.
Результат измерения зависит от характеристик измеряемого объекта,
взаимодействующего с ИП и искаженного измерением,
а исследователя, как правило, интересуют характеристики объекта,
не возмущённого измерением.
Их связь моделируется идеальным ИП,
заданным оператором $U \colon \mathcal{F} \to \mathcal{U}$,
на вход которого поступает тот же сигнал, что и на вход ИП $A$,
но на его выходе сигнал $U f$ равен
интересующей исследоватея характеристике объекта исследования.
Задача редукции состоит в нахождении оператора редукции $R_*$,
для которого $R_* \xi$ --- наиболее точная версия $U f$;
оператор редукции $R_*$ синтезируется в вычислительном преобразователе.
Если в \eqref{eqn:measurement-scheme} $f$ --- априори произвольный вектор,
$\nu$ --- случайный вектор, принимающий значения в $\mathcal{X}$,
имеющий математическое ожидание $\Expect \nu = 0$
и невырожденный ковариационный оператор
$\Sigma_{\nu}$: $\forall x \in \mathcal{X}$ $\Sigma_{\nu} x = \Expect \nu (x, \nu)$,
то линейный оператор редукции $R_* \colon \mathcal{X} \to \mathcal{U}$ определяется
как минимизирующий максимальную по $f$
среднеквадратичную (с. к.) погрешность интерпретации $R \xi$ как $U f$:
\begin{equation*}
h(R, U) = \sup_{f \in \mathcal{F}} \lVert R \xi - U f\rVert^2.
\end{equation*}
Эта погрешность минимальна \cite{pytyev_ivs} при
\begin{equation}
\label{eqn:lin-red-op}
R_* = U (A^* \Sigma_{\nu}^{-1} A)^{-1} A^* \Sigma_{\nu}^{-1}
\end{equation}
и равна
\begin{equation}
\label{eqn:lin-red-mse}
h(R_*, U) = \tr U (A^* \Sigma_{\nu}^{-1} A)^{-1} U^*,
\end{equation}
где ${}^-$ обозначает операцию псевдообращения,
если $U (I - A^- A) = 0$, и равна бесконечности в противном случае.

Пусть исследователя интересует распределение прозрачности объекта.
Значения прозрачностей пикселей принадлежат единичному отрезку.
Это учитывается при редукции измерения
проецированием оценки на множество $[0, 1]^{\dim \mathcal{U}}$
\cite{reduction_vmu, reduction_uzmu, ghost_images_jetp}:
оценка определяется как неподвижная точка отображения,
комбинирующего результат линейной редукции $R_* \xi$
и некоторую оценку $\hat{u}$ распределения прозрачности
как некоррелированные результаты <<основного>> и фиктивного измерений
с последующим проецированием на $[0, 1]^{\dim \mathcal{U}}$
при минимизации расстояния Махаланобиса $\lVert \Sigma_{R_* \xi}^{-1/2} \cdot \rVert$,
связанного с ковариационным оператором
$\Sigma_{R_* \xi} = U (A^* \Sigma_{\nu}^{-1} A)^{-1} U^*$
оценки линейной редукции.
Кроме того, исследователю известно, что
прозрачности соседних пикселей, как правило, отличаются слабо.
Эта информация часто формализуется
\cite{imaging_small_n_photons, gi_compressed_sensing_substr_const, hi-res_gi_sparsity, compressive_ghost_imaging, katz_et_al}
\emph{разреженностью} распределения прозрачности как вектора $U f$ в заданном базисе,
т.\,е. как информация, что в этом базисе
значительная часть компонент вектора $U f$ равна нулю.
В \cite{mgi_vmu, balakin_et_al_2019} предложен алгоритм редукции,
позволяющий исследователю учесть такую информацию
при обработке мультиплексированных квантовых фантомных изображений.
Алгоритм основан на проверке статистических гипотез
о равенстве компонент оценки в выбранном базисе нулю (альтернатива --- неравенство).
Его результат зависит от параметра алгоритма $\tau$~--–
уровня используемого в задаче проверки гипотез критерия.
Выбор этого параметра определяется приемлемым для исследователя компромиссом
между подавлением шума и искажением изображения.

\section{Влияние квантовой эффективности детекторов на схему измерения}

Пусть ПЗС-матрицы в объектном канале и восстанавливающем канале одинаковы и описываются
оператором $A_0$.
При единичной квантовой эффективности детекторов в объектном и восстанавливающем канале
схема измерения \eqref{eqn:measurement-scheme} принимает вид
\begin{equation*}
\xi =
\begin{pmatrix}
\xi_0\\
\xi_1
\end{pmatrix}
=
\begin{pmatrix}
A_0\\
A_0
\end{pmatrix}
n f
+
\begin{pmatrix}
\nu_{\textnormal{img}} + \nu_{\epsilon}\\
\nu_{\textnormal{img}}
\end{pmatrix},
\end{equation*}
где $\xi_{0}$~--- результат измерений ПЗС-матрицы в объектном канале,
$\xi_{1}$~--- результат измерения сформированного квантового фантомного изображения,
$\nu_{\textnormal{img}}$~--- часть погрешности, обусловленная регистрацией квантовых изображений,
$\nu_\epsilon$~--- часть погрешности квантового изображения, сформированного в объектном канале,
обусловленная шумовыми фотонами
(при формировании фантомного изображения они подавляются схемой совпадений,
если ее время срабатывания достаточно мал\'{о},
см. сравнение схем формирования квантовых изображений в \cite{imaging_small_n_photons};
в противном случае аналогичное слагаемое появляется
и в погрешности квантового фантомного изображения),
$n$~--- среднее число освещающих объект фотонов.
Ковариационный оператор погрешности имеет вид
\begin{equation}
\label{eqn:cov-op-unit-eff}
\Sigma_{\nu} =
\begin{pmatrix}
A_0 S(f) A_0^* + \Sigma_{\nu_{\epsilon}} & A_0 S(f) A_0^*\\
A_0 S(f) A_0^* & A_0 S(f) A_0^*
\end{pmatrix}
=
\begin{pmatrix}
1 & 1\\
1 & 1\\
\end{pmatrix}
\otimes
A_0 S(f) A_0^*
+
\begin{pmatrix}
\Sigma_{\nu_{\epsilon}} & 0\\
0 & 0
\end{pmatrix}
\end{equation}
где оператор $S(f)$ определяется дисперсиями и ковариациями фотоотсчетов.
Например, если статистика фотоотсчетов~--- пуассоновская,
то $S(f) = n \diag(f)$.
Здесь и далее $\diag(q)$~--- матрица,
диагональные элементы которой равны соответствующим компонентам $q$,
а остальные~--- нулю,
$\otimes$ обозначает произведение Кронекера.
В этом случае регистрация пространственного распределения интенсивности
излучения в объектном канале
не позволяет улучшить качество редукции
(результат этого измерения не влияет на результат редукции).

При эффективности детектора в объектном канале,
равной $\eta_0$,
и эффективности детектора в восстанавливающем канале,
равной $\eta_1$,
схема измерения принимает вид
\begin{equation*}
\xi =
\begin{pmatrix}
\xi_0\\
\xi_1
\end{pmatrix}
=
\begin{pmatrix}
\eta_0 A_0\\
\eta_0 \eta_1 A_0
\end{pmatrix}
n f
+
\begin{pmatrix}
\nu_{\textnormal{img}} + \nu_0 + \nu_{\epsilon}\\
\nu_{\textnormal{img}} + \nu_0 + \nu_1
\end{pmatrix},
\end{equation*}
где добавочные погрешности $\nu_0$ и $\nu_1$ обусловлены пропуском части фотонов
детекторами в объектном и восстанавливающем канале, соответственно.
Ковариационный оператор погрешности в этом случае имеет вид
\begin{multline}
\label{eqn:cov-op}
\Sigma_{\nu} =
\eta_0^2
\begin{pmatrix}
1 & \eta_1\\
\eta_1 & \eta_1^2
\end{pmatrix}
\otimes
A_0 S(f) A_0^*
+\\+
\eta_0^2
\begin{pmatrix}
\Sigma_{\nu_{\epsilon}} & 0\\
0 & 0
\end{pmatrix}
+
\begin{pmatrix}
\eta_0 (1 - \eta_0) & \eta_0 (1 - \eta_0) \eta_1\\
\eta_0 (1 - \eta_0) \eta_1 & \eta_0 \eta_1 (1 - \eta_0 \eta_1)
\end{pmatrix}
\otimes
\diag(A_0 f),
\end{multline}
где первые два слагаемых имеют то же происхождение,
что и в \eqref{eqn:cov-op-unit-eff},
но ослаблены из-за уменьшения среднего числа фотоотсчетов,
а последнее слагаемое непосредственно связано с неединичной эффективностью датчиков.

\begin{figure}
\centering
\includegraphics[scale=1]{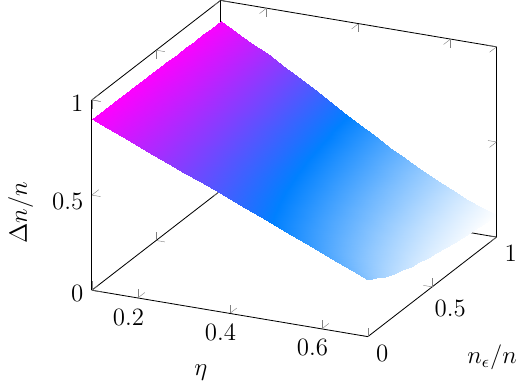}
\caption{Относительный выигрыш $\Delta n / n$ по среднему числу фотонов,
освещающих объект,
в зависимости от квантовой эффективности датчиков $\eta = \eta_0 = \eta_1$
и относительного среднего числа шумовых фотонов $n_\epsilon / n$
(по отношению к среднему числу $n$ фотонов освещения,
используемому при регистрации только фантомного изображения)}
\label{fig:mean-photon-number-gain}
\end{figure}

Поскольку оператор линейной редукции имеет вид \eqref{eqn:lin-red-op},
а погрешность оценки редукции при отсутствии дополнительной информации равна \eqref{eqn:lin-red-mse},
то выигрыш, обусловленный регистрацией изображения в объектном канале,
равен
\begin{multline}
\label{eqn:mse-gain}
\eta_0^{-2}
\tr
U
\left(
\eta_0 \eta_1^{-1}
\left(
A_0^* \left(
\eta_0 \eta_1 A_0 S(f) A_0^* + (1 - \eta_0 \eta_1) \diag(A_0 f)
\right)^{-1} A_0
\right)^{-1}
\right.
-\\-
\left.
\left(
\begin{pmatrix}
A_0^* & \eta_1 A_0^*
\end{pmatrix} \Sigma_{\nu}^{-1} \begin{pmatrix}
A_0\\
\eta_1 A_0
\end{pmatrix}
\right)^{-1}
\right)
U^*,
\end{multline}
в котором оператор $\Sigma_{\nu}$ определен выражением \eqref{eqn:cov-op}.
На рис.~\ref{fig:mean-photon-number-gain} показан относительный выигрыш
по среднему числу фотонов, освещающих объект,
в зависимости от квантовой эффективности датчиков
и относительного среднего числа шумовых фотонов
(по отношению к среднему числу фотонов освещения,
используемому при регистрации только фантомного изображения).
Выигрыш определен как отношение
уменьшения $\Delta n$ среднего числа фотонов, требуемого для освещения объекта,
к среднему числу $n$ фотонов освещения в традиционно применяемой схеме
при неизменной погрешности восстановленного изображения.
Предполагается, что при этом число шумовых фотонов пропорционально уменьшается.
Как и следовало ожидать, выигрыш монотонно возрастает
при уменьшении числа шумовых фотонов
и при уменьшении квантовой эффективности датчиков $\eta = \eta_0 = \eta_1$
(на графике предполагается,
что эффективность датчиков в объектном канале и в восстанавливающем канале одинакова).
При отсутствии шумовых фотонов относительный выигрыш равен $1 - \eta$.
При наличии дополнительной информации выигрыш, как правило, уменьшается
с сохранением знака,
так как одновременно уменьшаются обе сравниваемые погрешности,
причем б\'{о}льшая уменьшается в большей степени.

\section{Результаты компьютерного моделирования}

\begin{figure}
\centering
\begin{subfigure}[t]{\subfigsize}
\centering
\includegraphics[scale=1]{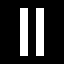}
\caption{Распределение прозрачности объекта исследования}
\label{fig:fig3_src}
\end{subfigure}
\begin{subfigure}[t]{\subfigsize}
\centering
\includegraphics[scale=1]{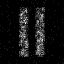}
\caption{Изображение, полученное матрицей в объектном канале}
\label{fig:fig3_obj}
\end{subfigure}
\begin{subfigure}[t]{\subfigsize}
\centering
\includegraphics[scale=1]{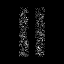}
\caption{Фантомное изображение (ФИ)}
\label{fig:fig3_ref}
\end{subfigure}
\begin{subfigure}[t]{\subfigsize}
\centering
\includegraphics[scale=1]{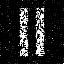}
\caption{Редукция без информации о разреженности}
\label{fig:fig3_red}
\end{subfigure}
\begin{subfigure}[t]{\subfigsize}
\centering
\includegraphics[scale=1]{{{fig3_red-0.1}}}
\caption{Редукция при информации о разреженности, $\tau = 0.1$}
\label{fig:fig3_red-0}
\end{subfigure}
\begin{subfigure}[t]{\subfigsize}
\centering
\includegraphics[scale=1]{{{fig3_red-0.2}}}
\caption{Редукция при информации о разреженности, $\tau = 0.2$}
\label{fig:fig3_red-1}
\end{subfigure}
\begin{subfigure}[t]{\subfigsize}
\centering
\includegraphics[scale=1]{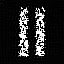}
\caption{Редукция ФИ без информации о разреженности}
\label{fig:fig3_red-s}
\end{subfigure}
\begin{subfigure}[t]{\subfigsize}
\centering
\includegraphics[scale=1]{{{fig3_red-0.1-s}}}
\caption{Редукция ФИ при информации о разреженности, $\tau = 0.1$}
\label{fig:fig3_red-0-s}
\end{subfigure}
\begin{subfigure}[t]{\subfigsize}
\centering
\includegraphics[scale=1]{{{fig3_red-0.2-s}}}
\caption{Редукция ФИ при информации о разреженности, $\tau = 0.2$}
\label{fig:fig3_red-1-s}
\end{subfigure}
\caption{Квантовые изображения объекта~(\subref{fig:fig3_src}),
зарегистрированные в (\subref{fig:fig3_obj})~объектном и (\subref{fig:fig3_ref})~восстанавливающем каналах
и результаты их обработки методом редукции:
обработка обоих изображений (\subref{fig:fig3_red})~без использования информации о разреженности
и (\subref{fig:fig3_red-0}, \subref{fig:fig3_red-1})~при наличии такой информации для двух значений параметра алгоритма $\tau$;
(\subref{fig:fig3_red-s}--\subref{fig:fig3_red-1-s})~обработка только фантомного изображения теми же методами}
\label{fig:fig3}
\end{figure}

На рис.~\ref{fig:fig3} показаны
результаты компьютерного моделирования и последующей обработки квантовых изображений
методом редукции измерения при помощи описанного в \cite{mgi_vmu, balakin_et_al_2019} алгоритма.
При моделировании предполагалось, что в обоих каналах расположены одинаковые матрицы,
датчики в которых имеют размер, втрое превышающий размер пикселя объекта,
каждый пиксель объекта освещается в средем $1$ фотоном,
среднее количество шумовых фотонов, не взаимодействующих с объектом,
но попадающий на матрицу в объектном канале~--- $0.1$ фотон на пиксель объекта,
а квантовые эффективности датчиков в объектном канале и в восстанавливающем канале
равны $0.4$.

Как отмечалось выше, параметр алгоритма $\tau$
отражает приемлемый для исследователя компромисс
между подавлением шума и искажением изображения.
Чем больше значение $\tau$, тем в большей степени подавляется шум,
но при этом также постепенно усиливается искажение деталей изображения
(ср., например, рис.~\ref{fig:fig3_red-0-s} и \ref{fig:fig3_red-1-s},
где при увеличении $\tau$ происходит размытие изображения щели).
Значение $\tau = 0$ соответствует
как отсутствию связанного с использованием информации о разреженности распределения прозрачности дополнительного подавления шума, так и неискажению изображения,
т.\,е. оно эквивалентно отсутствию информации
о разреженности распределения прозрачности объекта.
Значения $\tau$, использованные на рис.~\ref{fig:fig3},
выбраны так, чтобы на рис.~\ref{fig:fig3_red-0} искажения были бы еще не заметны,
а на рис.~\ref{fig:fig3_red-1}~--- уже заметны.

Высказанного выше предположение,
что при наличии дополнительной информации выигрыш за счет обработки изображения,
зарегистрированного в объектном канале, уменьшается
при сохранении знака,
подтверждается сравнением пар рис.~\ref{fig:fig3_red} и \subref{fig:fig3_red-s}
и рис.~\ref{fig:fig3_red-0} и \subref{fig:fig3_red-0-s}:
первые два отличаются больше, чем последние два,
поскольку сужается множество возможных значений оценки.

Уровень шума оценки, построенной только по фантомному изображению,
больше, чем построенной по обоим изображениям.
Поэтому размытие при одинаковых значениях параметра алгоритма $\tau$,
отвечающего за баланс между размытием важных деталей изображения и подавлением шума,
больше для оценки, построенной только по фантомному изображению.

Наконец, локализация шума,
обусловленного использованием полученного в объектном канале изображения,
отличается от локализации шума, связанного с неединичной эффективностью детекторов.
В то время, как распределение по площади изображения
шума, связанного с шумовыми фотонами, как правило, не зависит от объекта,
распределение по площади изображения
шума, связанного с неэффективностью детекторов
напрямую от него зависит, так как чем больше фотонов попадает на датчик,
тем больше дисперсия числа его срабатываний.
Поэтому изменение параметра $\tau$ по-разному влияет на искажение деталей изображения,
расположенных в областях с различной средней яркостью.

\section*{Заключение}

В заключение отметим, что рассмотренная схема позволяет улучшить качество квантового изображения
на этапе компьютерной обработке также
за счет учета на этом этапе дифракционной ограниченности фантомных изображений \cite{belinsky_2018, moreau_et_al_2018, moreau_et_al_2018a}.
Это связано с тем, что
обычное изображение не имеет дифракционных ограничений,
характерных для фантомных изображений,
связанных с ограниченностью поперечных размеров накачки и,
соответственно, ее расходимостью.

Хотелось бы отметить еще одно интересное схемное решение,
базирующееся на предложенном варианте формирования квантовых фантомных изображений:
одновременное изучение двух объектов, каждый из которых расположен в своем канале.
При этом мы получаем 2 пары квантовых изображений,
2 обычных и 2 фантомных, за один сеанс,
ценой ухудшения качества по сравнению с отсутствием второго объета.
Ухудшение связано с тем, что второй объект не пропускает часть фотонов,
коррелированных с фотонами в канале первого объекта,
и наоборот.
Тем не менее, в некоторых ситуациях это могло бы быть полезным.

Работа выполнена при финансовой поддержке РФФИ (грант №18-01-00598 А).

\bibliographystyle{ugost2008}
\newcommand{\BibDash}{}
\bibliography{gi-impr-obj}